\documentclass[%
 reprint,
superscriptaddress,
nofootinbib,
 amsmath,amssymb,
 aps,
]{revtex4-2}

\usepackage{graphicx}
\usepackage{dcolumn}
\usepackage{bm}
\usepackage{hyperref}

\usepackage{xcolor}
\usepackage{soul}
\usepackage{placeins}
\usepackage{xr-hyper}

\begin{document}

\preprint{APS/123-QED}

\title{Recurring region for neutron-star observables}

\author{Alexander Clevinger}
 \affiliation{Center for Nuclear Research, Department of Physics, Kent State University, Kent, OH, USA 44242}
\author{Zidu Lin}%
 \affiliation{Department of Physics and Astronomy, University of Tennessee Knoxville, Knoxville, TN 37996-1200, USA.}
 \affiliation{Department of Physics and Astronomy, University of New Hampshire, Durham, NH 03824, USA}

\author{Milena Albino}
\affiliation{Department of Physics, CFisUC, University of Coimbra, P-3004 - 516 Coimbra, Portugal}

\author{Peter Hammond}
\affiliation{Department of Physics, The Pennsylvania State University, University Park, PA 16802, USA}
\affiliation{Institute for Gravitation and the Cosmos, The Pennsylvania State University, University Park, PA 16802, USA}
\affiliation{Department of Physics and Astronomy, University of New Hampshire, Durham, NH 03824, USA}

\author{Andrew Steiner}
\affiliation{Department of Physics and Astronomy, University of Tennessee Knoxville, Knoxville, TN 37996-1200, USA.}

\affiliation{Physics Division, Oak Ridge National Laboratory, Oak Ridge, TN 37831, USA}

\author{Veronica Dexheimer}
 \affiliation{Center for Nuclear Research, Department of Physics, Kent State University, Kent, OH, USA 44242}

\date{\today}

\begin{abstract}
In this letter, we report a novel, somewhat analytical way to produce equations of state (EOSs) that generate particular values of neutron star mass, radius,  and tidal deformability. This is possible because our description for the EoS of dense matter can produce recurring regions, small areas where several EoSs cross in the mass-radius and mass-tidal deformability diagrams. We can place recurring regions in desired locations of these diagrams, corresponding e.g., to a given observation. Our prescription is versatile, in the sense that different microscopic models can be used for the low density hadronic phase and high density quark phase, as long as they are connected by a percolation, a description that mimics quark deconfinement and is motivated by QCD. 
The several EoSs that pass by a recurring region can present different thresholds for the boundaries of the percolation region (different beginning and ending for the quark deconfinement region), as well as different orders for the phase transition at the boundaries. When combining all these features, our prescription allows one not only to produce an EoS that matches an observation, but also one that matches specific chosen criteria for the EoS. The EoSs produced by this new method will be specially suitable for the study of dense-matter properties in future gravitational-wave observations, when both the inspiral and post-merger phase signals will become available. Our numerical code that calculates recurring regions using CompOSE microscopic EoSs is open source and publicly available.

\end{abstract}

\maketitle


\section{Introduction}
\label{sec:intro}

Within the high densities achieved in the interior of neutron stars, we expect exotic matter to appear within their cores~\cite{Baym:2017whm}. This comprehends, for example, hadronic matter containing (besides protons and neutrons) hyperons and Delta baryons and a quark phase containing up, down, and strange quarks (see the Supplemental Materials (SM) \cite{appendixA} for a quantitative motivation for quark deconfinement). With recent exponential growth of neutron-star observations through direct observation of light, as well as gravitational wave observation from their mergers, there is a robust need to build up realistic models for dense matter, while also keeping them consistent with findings from terrestrial laboratory experiments. Exotic matter, particularly strongly interacting deconfined quark matter, is poorly understood and neutron stars can provide us with further insight into the low-temperature, high-density regime of the Quantum Chromodynamics (QCD) phase diagram (see Fig.~1 from~\cite{MUSES:2023hyz}). This means that the deconfinement phase transition could be observed inside neutron stars or during their mergers in the near future~\cite{Most:2018eaw,Bauswein:2018bma}, pointing toward the natural (not created in the laboratory) occurrence of deconfined quark matter, as long as we know how to interpret it as such.

In this letter we work on providing such interpretation for deconfinement.
Whereas strong (or first-order) deconfinement phase transitions generally cause kinks and more complicated structures in mass-radius diagrams for neutron stars \cite{Alford:2013aca}, smoother (or higher-order) phase transitions do not. We focus on the latter and build EoSs using a quantum chromodynamics (QCD) motivated description referred to as percolation~\cite{Kojo:2014rca} that generates bump structures in the speed of sound of matter. Such bumps have been motivated by current constraints of nuclear physics and astrophysics~\cite{Tan:2021nat,Jakobus_2021,Dexheimer_2015,Alford_2017,Zacchi_2016,Alvarez_Castillo_2019,Li_2020,wang2020exploringhybridequationstate,Fadafan_2020,Blaschke_2020,Dutra_2016,McLerran_2019,Xia_2021,Yazdizadeh_2019,Shahrbaf_2020,Zhao_2020,Lopes_2021,Duarte:2020xsp,rho2021fractionalizedquasiparticlesdensebaryonic,Marczenko_2020,Minamikawa_2021,Hippert:2021gfs,Pisarski:2021aoz,Stone:2019blq,Ferreira_2020,Kapusta_2021,Somasundaram_2022,kojo2021qcdequationsstatespeed,Mukherjee_2017,Li_2020_2,Jin_2022,Lee_2022,Raduta_2021,Dutra_2012,Dexheimer:2009hi,Dexheimer:2008ax,Guichon_1996,Malfatti_2020,Bedaque:2014sqa,Annala_2020,Stone_2021,Gusakov_2014,de_Paoli_2013,PhysRevC.103.045804,Tu_2022,Monta_a_2019,Kojo_2019,Malfatti_2019,Jokela_2021,Jeong_2020,Sen_2021_2,Motornenko:2019arp,Baym:2019iky,Li_2018,kumar2023nonradialoscillationmodeshybrid,Marczenko_2022,Tews:2018kmu,Fujimoto:2022ohj,Zuo_2022,braun2022zerotemperaturethermodynamicsdenseasymmetric,Ivanytskyi_2022,fraga2022strangequarkmatterbaryonic,Huang_2022,Han:2022rug,Pinto_2023,kumar2023quarkyonicmodelneutronstar,Liu_2023,yamamoto2023quarkphasesneutronstars,kouno2024hadronquarktransitionchiralsymmetry,gao2024reconcilinghessj1731347constraints,tajima2024triplingfluctuationspeakedsound,Legred_2022,Altiparmak_2022,mroczek2023nontrivialfeaturesspeedsound,Kawaguchi_2024,Cuceu_2025,xia2024astrophysicalconstraintsnucleareoss,ye2025highdensitysymmetryenergy,Kov_cs_2022}. 
We analyze different microscopic EoSs for the hadronic and quark phases, as well as different parameter sets for the percolation description, including different thresholds for the boundaries of the percolation region and different orders for the phase transition at the boundaries (second and third). We provide a new reliable method based on the average speed of sound inside a neutron star to identify which percolation parameters result in EoSs that generate a stellar sequence passing by one small recurring region in the mass-radius and mass-tidal deformability diagrams. We also show that this recurring region also occurs in chirp mass-tidal deformability space (independent of mass ratio), which can be measured more directly in gravitational wave observations.

We begin by discussing fixed stellar masses (that e.g., matches an observation), and then discuss how to change the respective radii or tidal deformability of the recurring region. We then discuss how this universal behavior can be used to interpret future observations of neutron stars and neutron-star mergers, e.g., how a given mass and radius or tidal deformability observation can be described by EoSs with different stiffness and/or quark deconfinement threshold. Note that merger events of neutron-stars residing in recurring regions typically have non-distinguishable inspiral-phase gravitational-wave signals but present very different post-merger phase signals, which is discussed in our follow-up work~\cite{Hammond:2025kki}. The numerical code used in our work is open source and publicly available~\cite{github}.

\section{Methodology}
\label{Formalism}
To describe $\beta$-equilibrated, charge-neutral, dense zero-temperature hadronic matter at lower densities, we make use of two realistic microscopic EoS models. First, we consider the Chiral Mean Field (CMF) model, an effective relativistic model which was constructed and fitted to reproduce the main features of QCD, nuclear saturation properties, and  neutron-star observables~\cite{Dexheimer:2008ax}. We focus on one parametrization from \cite{Clevinger:2022xzl} available on the CompOSE repository~\cite{Oertel:2016bki,Typel:2013rza,Oertel:2016bki,CompOSECoreTeam:2022ddl,Typel:2013rza}, referred to as CMF-7. It includes nucleons, hyperons, Deltas baryons, electrons and muons, combined with additional  vector-isovector interactions and is connected to a crust EoS for the lowest densities to describe nuclei~\cite{Gulminelli:2015csa}. The second hadronic EoS is DD2F, a relativistic mean field (RMF) approach with density-dependent couplings and corrections due to flow constraint~\cite{Alvarez-Castillo:2016oln}. It includes nucleons and a statistical model for the crust~\cite{Hempel:2009mc}.

To describe $\beta$-equilibrated, charge-neutral, dense zero-temperature quark matter at higher densities, we make use of two microscopic prescriptions that are already connected to the hadronic EoSs we selected. Ref.~\cite{Clevinger:2022xzl} already provides a hybrid version of CMF-7 connected to a quark phase containing up, down, and strange quarks (in addition to leptons) through a first-order phase transition~\cite{Dexheimer:2009hi,Dexheimer:2020rlp}. \cite{Bastian_2021} already connects the DD2F hadronic phase with a phase of up and down quarks described by the The Nambu-Jona-Lasinio (NJL) model used with a string flip model (SFM) term and vector interactions~\cite{PhysRevD.96.056024}. Notice that both,  the CMF and NJL models, additionally describe the restoration of chiral symmetry, predicted by QCD to take place at high energies~\cite{Nambu:1961tp}.

To describe the \textit{process} of deconfinement, we insert a percolation around the first-order phase transition in the different combined hadronic and quark EoSs. We follow the  method described in~\cite{Kojo:2014rca}, which removes the discontinuities associated with the first-order phase transition. The percolation introduces instead two new phase transitions (now of higher order) at the boundaries of a polynomial, whose coefficients are fitted to various boundary conditions. In this work, we use fifth order polynomials to describe the pressure around quark deconfinement.

The idea of percolation was motivated by the quarkyonic picture, based on the idea that between the hadronic and quark phases there is a phase where the baryon structure begins to break down, but the quarks themselves are still somewhat bound to each other~\cite{McLerran:2020rnw}. In the limit that hadrons are made up of quarks with a large number of colors, this is a prediction of QCD. In our Universe, where hadrons are made up of quarks with three colors, quarkyonic matter might still persist and, at low temperatures, such phase could span a large range of densities, corresponding to a large portion of neutron star interiors. The percolation method allows us to produce a middle ground that  thermodynamically provides a region that resembles a quarkyonic phase~\cite{McLerran:2007qj}. This comes at the expense of losing the microphysics from the original EoSs, including particle densities and chemical potentials, inside the percolation.

\begin{figure*}[t!]
\centering 
\includegraphics[width=1\textwidth,angle=0]{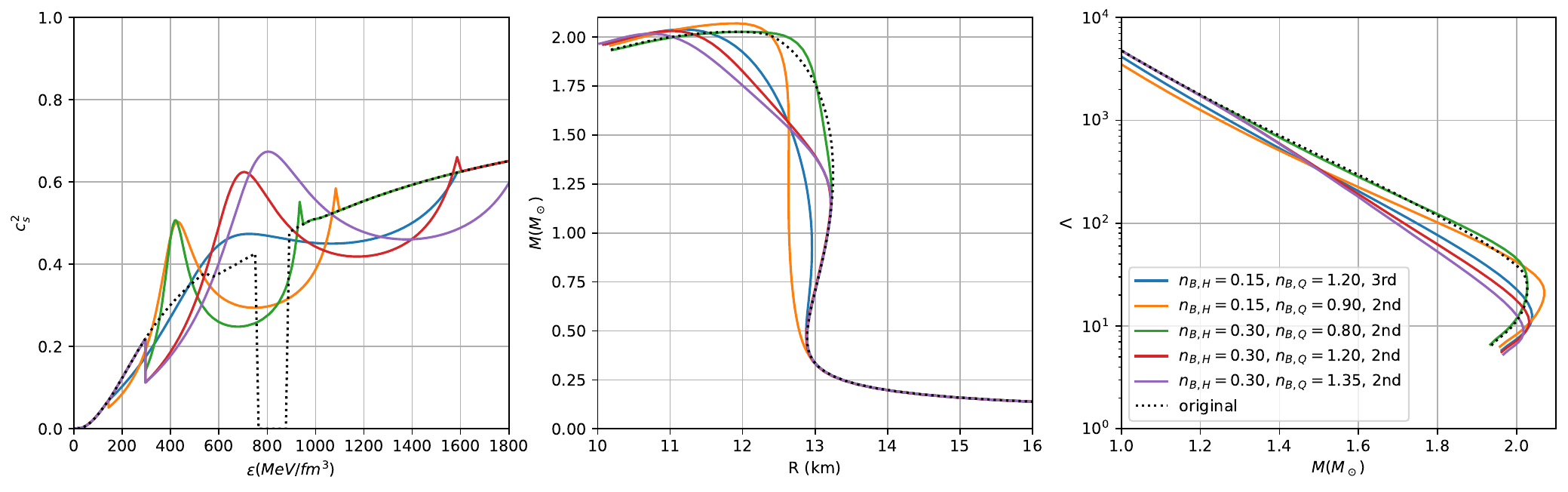}	
\caption{Speed of sound for five CMF-7 EoSs with percolation and original microscopic CMF-7 EoS (left panel), corresponding mass-radius (middle panel), and mass-tidal deformability diagrams (right panel). The EoSs with percolation are built using different densities (in fm$^{-3}$) and reproducing different orders for the phase transition at the hadronic and quark boundaries.} 
\label{tov}
\end{figure*}

Our procedure to build the percolation is as follows: 1) we start by choosing for each EoS two values of baryon (number) density $n_B$ to control the beginning and end of the percolation region, $n_{B,H}$ and $n_{B,Q}$, 2) we find the corresponding baryon chemical potentials $\mu_{B_H}$ and $\mu_{B_Q}$ in the hadronic and quark EoSs, and 3) we calculate the pressure in the percolation region using the polynomial,
%
    $\mathcal{P} = \sum_{m = 0}^{5} b_m (\mu_B)^m,$
%
with 6 coefficients determined by 6 boundary conditions (that determine the new order of phase transitions at the boundaries)
\begin{eqnarray}
  \frac{d^NP_H}{d\mu_B^N}\Bigr|_{\mu_{BH}}= \frac{d^N\mathcal{P}}{d\mu_B^N}\Bigr|_{\mu_{BH}},\quad  \frac{d^NP_Q}{d\mu_B^N}\Bigr|_{\mu_{BQ}}= \frac{d^N\mathcal{P}}{d\mu_B^N}\Bigr|_{\mu_{BQ}},
    \label{boundaries}
\end{eqnarray}
where for $N=0$ these mean that the percolation region pressure matches the EoS pressure on either side (applies even to a first-order phase transition), for $N=1$ these mean that the first derivatives match (second-order phase transition and higher), and for $N=2$ these mean that the second derivatives match (third-order phase transition and higher). We note that the $\rm{n}^{\rm{th}}$ derivative of the pressure can also be written as the $\rm{n}^{\rm{th}}$-order susceptibility, 
$\chi_n^B=\frac{\partial^n P}{\partial \mu_B^n}=\frac{\partial^{n-1} n_B}{\partial \mu_B^{n-1}}$.

Alternatively, one can also impose specific constraints (beyond Eq.~\ref{boundaries}) at the boundaries, such as specifying any of the susceptibilities $\chi_n^B$. This allows us to change the size of the discontinuities at the boundaries of the percolation region. In this work, we do exactly that by:
%
{\bf{i)}} matching 0th- and 1st-order derivatives at the boundaries and varying $\chi_2^B$ on both sides of the percolation region. This leads to a fifth-order polynomial to describe a {\bf{second-order phase transition}}. The 4 constraints are fixed by Eq.~\ref{boundaries}) (with $N=0$ and $N=1$)
{\bf{ii)}} matching 0th-, 1st-, and 2nd-order derivatives at the boundaries. This leads to a fifth-order polynomial to describe a {\bf{third-order phase transition}}. The 6 constraints are fixed by Eq.~\ref{boundaries}) (with $N=0$, $N=1$, and $N=2$).
%
The energy density $\varepsilon$ in the percolation region is calculated as $\varepsilon=-\mathcal{P}+\mu_B~ \partial \mathcal{P}/\partial \mu_B$.

Finally, we must also consider whether our newly constructed EoSs are physical. There are several thermodynamic and astrophysical constraints to consider, in addition to causality (speed of sound less than the speed of light, $1$ in natural units). Other constraints at low or high density are already fulfilled by the original microscopic EoSs. We impose a limit within the percolation region on the speed of sound ($0< c_s^2 < 1$), that the pressure must be continuous, and that $n_B$ must be a convex function of $\mu_B$, $\chi_2^{B}\ge 0$ (see 
\cite{Aloy:2018jov} for a discussion of concave EoSs in the context of gravitational waves). The main astrophysical constraint we impose is that the EoS must be able to produce a neutron star with mass $>2\,\rm{M}_{\odot}$ to be consistent with astrophysical observations of heavy neutron stars~\cite{Fonseca:2021wxt}. This can be checked by solving the Tolman-Oppenheimer-Volkoff (TOV) equations  with our constructed EoSs~\cite{Tolman:1939jz,Oppenheimer:1939ne}. Other relevant astrophysical observables, such as radius or tidal deformability (which requires solving Einstein field equations with perturbations due to the gravitational field~\cite{Hinderer:2007mb}), can also put constraints on producing physical EoSs. However, since the uncertainties on those measurements are much larger, we only use them as a guide for the discussion of hybrid stars built from EoSs with percolation, more specifically reproducing stellar radii that are not too large. For further discussion on these constraints, see~\cite{MUSES:2023hyz}.

\section{Results}
\label{Observables}

Fig.~\ref{tov} shows results for
several EoSs generated from CMF-7 using our percolation prescription and enforcing the constraints listed above together with small neutron star radii, below $\sim 13.5$ km~\cite{Miller:2019cac,Riley:2019yda}. They are built using different densities on the hadronic and quark percolation boundaries, reproducing different orders for the phase transitions at these boundaries, and imposing different susceptibilities at the boundaries for the case of second-order phase transitions. As seen in the left panel, they produce bumps in the speed of sound squared. For the original microscopic CMF-7 EoS, the speed of sound goes to zero, marking the deconfinement first-order phase transition.  The middle panel shows the mass-radius diagram and the right panel shows the mass-tidal deformability diagram for these curves. Both diagrams present a recurring region, where several different  EoSs with percolation (but not all) cross at a stellar mass $\sim1.5\,\rm{M}_{\odot}$. 

\subsection{Defining a recurring region}

\begin{figure*}[t!]
\centering 
\includegraphics[width=.65\textwidth, angle=0]{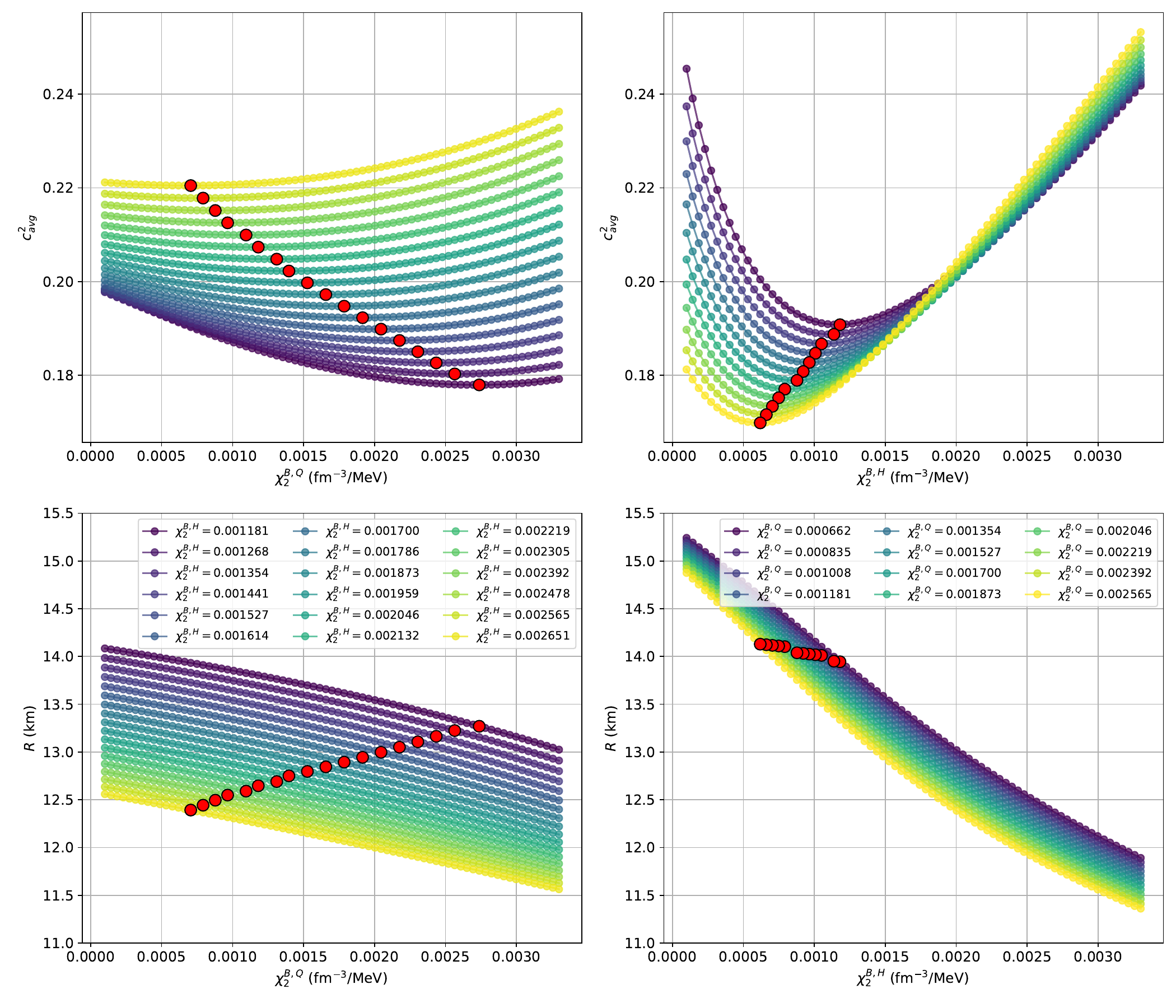}	
\caption{Average speed of sound squared for different second-order susceptibilities at the percolation boundaries using the same density boundaries as the blue curve of Fig.~\ref{tov} for CMF-7 for a star with mass $1.554\,\rm{M}_{\odot}$ (top panels). 
The left panel shows the susceptibility at the quark boundary in the x-axis and the different susceptibilities at the hadronic boundary in different colors (values in fm${^{-3}}$/MeV). The right panel shows the susceptibilities exchanged. Red dots mark the minimum $c_\mathrm{avg}^2$ for each curve. The lower panels show the corresponding stellar radii.}
\label{v1_fig}
\end{figure*}

To investigate this recurring region, we define a new quantity, the average speed of sound squared,
\begin{eqnarray}
    c^2_\mathrm{avg} = \frac{P_c-P_H}{\varepsilon_c-\varepsilon_H}\,,
    \label{percolation2}
\end{eqnarray}
which measures the average speed of sound within the percolation region up to the center of the chosen star. In the SM, Ref \cite{appendixB1} discusses why this quantity is the relevant one for our discussion. Note that the average speed of sound squared in the entire star has been recently connected to macroscopic stellar properties, such as compactness, through universal relations~\cite{Saes:2024xmv}. 
%
As an example of our prescription, we calculate the average speed of sound from the beginning of the percolation region up to the center of the star for the first EoS with percolation from Fig.~\ref{tov} (blue curve), corresponding to a third-order phase transition with boundaries $n_{B,H}=0.15$ fm$^{-3}$ and $n_{B,Q}=1.20$ fm$^{-3}$, using a stellar mass of $1.554\,\rm{M}_{\odot}$ (where the recurring region appears in Fig.~\ref{tov}). Then, we modify the percolation by varying widely two constraints, the susceptibilities at the boundaries, $\chi_2^{B,H}$ and $\chi_2^{B,Q}$, (turning them into second-order phase transitions) and compare the $c^2_\mathrm{avg}$ from Eq.~\ref{percolation2} across a large parameter space. The results for the causal and convex EoSs are shown in the top panels of Fig.~\ref{v1_fig}, where the minima in $c^2_\mathrm{avg}$ are marked by red dots.

More specifically, the top left panel of Fig.~\ref{v1_fig} shows the susceptibility at the quark boundary for the percolation in the x-axis and different values for the susceptibility at the hadronic boundary in different colors. The top right panel shows instead the susceptibility at the hadronic boundary for the percolation in the x-axis and different values for the susceptibility at the quark boundary in different colors. The lower panels of Fig.~\ref{v1_fig} show the corresponding stellar radii. In the bottom left panel, the radii corresponding to the minima marked by the red dots vary in a small range of $\sim$ 0.8 km and in the right panel in an even smaller range of $\sim$ 0.2 km. Nevertheless, the stellar radii in the right panel correspond to values that are too large, outside the current accepted values for astrophysics (see Tab.~10 of~\cite{MUSES:2023hyz}), and, therefore, were not selected for Fig.~\ref{tov}. The stellar radii range in the bottom left panel of Fig.~\ref{v1_fig} corresponds to values that are smaller, compatible with astrophysics data, and include the values for the first (blue) curve of Fig.~\ref{tov} ($\chi_2^{B,H}= 0.00226$ and $\chi_2^{B,Q} = 0.00106$ fm${^{-3}}$/MeV), as expected. Furthermore, the y-axis range corresponding to the red dots in the bottom left panel of Fig.~\ref{v1_fig} provides a size for the recurring region in Fig.~\ref{tov} (for a given set of density boundaries for the percolation).

\subsection{Applying recurring regions to different cases}

To verify the robustness of our prescription, we repeat our calculations for the CMF-7 EoS producing a recurring region for different cases. We start with a percolation constructed using different density boundaries for CMF-7 (that match the red curve of Fig.~\ref{tov}), still for a star with mass $1.5\,\rm{M}_{\odot}$. 
We once more vary the susceptibility at the hadronic and quark boundaries. See the SM ~\cite{appendixB2} for details. The results are similar to the ones of Fig.~\ref{v1_fig}. The procedure shown in the left panels of Fig.~\ref{v1_fig} now produces no causal EoSs that are minima, while the procedure shown in the right panels of Fig.~\ref{v1_fig} now produces roughly the same recurring region as in Fig.~\ref{v1_fig}. 

We also repeat our calculations for the CMF-7 EoS producing a
recurring region for a larger mass of $1.8\,\rm{M}_{\odot}$. We report
the details in the SM~\cite{appendixB3}. We find again two recurring regions, but exclude the large radius one due to astrophysical constraints. Still both recurring regions appear at larger
radius than for the lower star mass. We also repeat our
calculations for different hadronic and quark microscopic
EoSs, the DD2F-NJL model combination. We report the details in the SM~\cite{appendixB4}. We choose a star with mass of $1.554\,\rm{M}_{\odot}$ and find again two recurring regions, as in Fig.~\ref{v1_fig}, but exclude the large radius one due to astrophysical constraints.

Now that we understand recurring regions,
we can return to discussing Fig.~\ref{tov}.
The blue, orange, red, and purple curves have susceptibilities
that are close to minima in $c_s^2$ for a $1.554\,\rm{M}_{\odot}$ star.
As a result, these curves pass close enough to describe
a recurring region.
This is the case even though they have different values of $n_{B,H}$ and $n_{B,Q}$ and different orders of phase transition.
The red and purple curves have higher $n_{B,H}$ values than the blue and orange curves.
This modifies their behavior for low densities (left panel of Fig.~\ref{tov})
and low-mass stars (middle and right panels of Fig.~\ref{tov}).
However, again, since their susceptibilities are close to the minima in $c_s^2$,
the curves pass by the recurring region.
The green curve, however, has susceptibilities that are NOT close to the minima
in $c_s^2$ and, therefore, does not pass close to the recurrent region.

\subsection{Implications of recurring regions}

Now that we can produce recurring regions for a given stellar mass, we can return to the discussion of how to obtain EoSs that match a real measurement. But, first, we discuss how to change the radius/tidal deformability of the recurring region. This can be done by changing the hadronic EoS used (changing the quark EoS would only affect significantly very high mass recurring regions). This can be done, e.g. by using the DD2F-NJL model combination, as discussed above. For the same stellar mass of $1.554\,\rm{M}_{\odot}$ star, going from CMF-7 in Fig.~\ref{v1_fig} to DD2F-NJL in Fig.~\ref{NJL} reduces the (midpoint) recurring region radius by $\sim 2.3 \%$ and also reduces the range of the recurring region by $\sim 0.3$ km. Changing the radius is then expected to change the tidal deformability, as these are related quantities~\cite{De:2018uhw,Raithel:2018ncd}). For a more consistent change of the recurring region radius, one could use e.g., different versions of the CMF and DD2 models, including different particles and different interactions. In particular, vector-isovector interactions have been shown to additionally modify the relation between stellar radius and tidal deformability. As a result, one could find mass-radius recurring regions for different CMF versions (with different vector-isovector interactions) and study how strong-force vector interactions change the associated tidal deformability. See \cite{Clevinger:2022xzl} for the description of a large variety of CMF EoSs, all available in CompOSE.

Once a recurring region is built, either in a mass-radius or mass-tidal deformability diagram, one has a set of EoSs, each with a somehow different characterization for quark deconfinement (compare blue, orange, red, and purple curves in the left panel of Fig.~\ref{tov}) that match a given observation. Since currently observations of neutron stars and neutron star mergers still provide somehow large intervals for mass-radius or mass-tidal deformability regions, midpoints could be used as input for our prescription. In the SM ~\cite{appendixC}, we also plot different quantities that are more closely related to neutron mergers and show that the recurring regions translate to this parameter space. They show that even in unequal-mass binaries, the recurring region still persists. See the SM ~\cite{appendixD} for a summary of the steps necessary to follow our prescription. 

Note that \textit{simply varying the parameters within one microscopic model does not guarantee to generate EoSs passing by a given region in mass-radius or mass-tidal deformability diagram. In addition, varying the parameters within one microscopic model does not generate a collection of EoSs passing by one small region.} One exception to the latter statement (but not the former) is the special point in the mass-radius diagram found in \cite{Cierniak:2020eyh,Gartlein:2023vif}, which was found by varying the onset of the first-order phase transition to deconfined quark matter in microscopic models. But, as this special point is a result of the combination of hadronic and quark microscopic models (without the intermediate percolation region), it cannot be placed at will in the mass-radius (or other) diagram. At the deconfinement first-order phase transition, the speed of sound goes to zero, as opposed to our case, in which the speed of sound presents a bump in the percolation region.

\section{Conclusions and Outlook}
\label{Conclusions}

In this work, we provide a new somewhat analytical prescription to produce realistic dense matter equations of state (EoSs) that generate stellar sequences in mass-radius and mass-tidal deformability diagrams that cross through a small recurring region. Our prescription can be used to produce a recurring region that matches a real observation of neutron star mass and radius or a neutron star merger observation of mass and tidal deformability by varying the quark deconfinement part of the EoS. Quark deconfinement is described by a percolation, a QCD motivated technique based on the quarkyonic picture. Our EoSs are realistic due to using microscopic hadronic and quark EoSs for low and high densities, respectively, together with a percolation that generates bump structures in the speed of sound. These bumps have been associated with e.g., higher-order vector interactions, new hadronic degrees of freedom, and symmetry restoration/breaking, in addition to deconfinement \cite{Jakobus_2021,Dexheimer_2015,Alford_2017,Zacchi_2016,Alvarez_Castillo_2019,Li_2020,wang2020exploringhybridequationstate,Fadafan_2020,Blaschke_2020,Dutra_2016,McLerran_2019,Xia_2021,Yazdizadeh_2019,Shahrbaf_2020,Zhao_2020,Lopes_2021,Duarte:2020xsp,rho2021fractionalizedquasiparticlesdensebaryonic,Marczenko_2020,Minamikawa_2021,Hippert:2021gfs,Pisarski:2021aoz,Stone:2019blq,Ferreira_2020,Kapusta_2021,Somasundaram_2022,kojo2021qcdequationsstatespeed,Mukherjee_2017,Li_2020_2,Jin_2022,Dexheimer:2009hi,Dexheimer:2008ax,Malfatti_2020,Bedaque:2014sqa,Annala_2020,Stone_2021,Gusakov_2014,de_Paoli_2013,PhysRevC.103.045804,Tu_2022,Monta_a_2019,Kojo_2019,Malfatti_2019,Jokela_2021,Jeong_2020,Sen_2021_2,Motornenko:2019arp,Baym:2019iky,Li_2018,kumar2023nonradialoscillationmodeshybrid,Marczenko_2022,Zuo_2022,braun2022zerotemperaturethermodynamicsdenseasymmetric,Ivanytskyi_2022,fraga2022strangequarkmatterbaryonic,Huang_2022,Han:2022rug,Pinto_2023,kumar2023quarkyonicmodelneutronstar,Liu_2023,yamamoto2023quarkphasesneutronstars,kouno2024hadronquarktransitionchiralsymmetry,gao2024reconcilinghessj1731347constraints,tajima2024triplingfluctuationspeakedsound,Kawaguchi_2024,Cuceu_2025,ye2025highdensitysymmetryenergy,Kov_cs_2022}. This is in contrast of e.g., using deep neural networks to perform an inversion of TOV back to the EoS \cite{Ventagli:2024xsh} (with further loss of microscopic information).

Once a recurring region is found, with several EoSs crossing it, these EoSs can be used to explore the consequences of the observation: how stiff the EoS can be, does it cross the conformal limit, what particles/interactions can it have, etc. We give a few examples of recurring points and related EoSs (see e.g., blue, orange, red, and purple lines in Fig.~\ref{tov}). Our work accompanies an open source script, that allows users to reproduce our work or create recurring regions and generate EOSs that match past and future observations.

At the time being, neutron star and neutron star merger observations still provide significant ranges for mass-radius and mass-tidal deformability measurements, but once observations become more accurate and/or we collect more data, it is expected that these regions will shrink. Note that recurring regions also appear in chirp mass-binary tidal deformability space for different mass ratios.
The lower mass recurring region we discuss in this letter is consistent with $95$ confidence regions in the mass-radius measurements from NICER and LIGO (see Fig.~31 in \cite{MUSES:2023hyz}). Going beyond tidal-deformability observations from neutron-star merger inspirals, recurring regions are also be interesting to decode post-merger gravitational wave signals. Neutron-star merger simulations performed using different EoSs with similar macroscopic properties (mass, radius, and tidal deformability) allow us to make comparisons only based on the EoS structure, see our recent work \cite{Hammond:2025kki} for more details.

Finally, note that this is an exploratory study. In the future, one could explore in a more consistent matter the size of the recurring region by using many more microscopic EoSs. This would require a more robust TOV solver (e.g., the one available by the MUSES collaboration~\cite{ReinkePelicer:2025vuh}), as the interpolation error in the mass-radius curves can propagate into the speed of sound average, and a statistical approach such as Bayesian analysis. 

\begin{acknowledgements}
This work is partially supported by the NP3M Focused Research Hub supported by the National Science Foundation (NSF) under grant No. PHY-2116686.  VD also acknowledges support from the Fulbright U.S. Scholar Program and the Department of Energy under grant DE-SC0024700. ZL and AWS were supported by NSFPHY21-16686. AWS was also supported by the Department of Energy Office of Nuclear Physics. MA expresses sincere gratitude to the FCT for their generous support through Ph.D. grant number 2022.11685.BD (DOI: \url{https://doi.org/10.54499/2022.11685.BD}. Finally, we thank the referees for the suggestion to explore quantities related to gravitational wave observations as we could see recurring regions in this parameter space as well. With this, we also thank Nicol\'as Yunes and Paul Lasky for helping us with the referees'  suggestion.

\end{acknowledgements}

\nocite{Shapiro:1983du,weber2007neutronstarinteriorsequation,Schaffner-Bielich_2020,alma991063201645604801,ParticleDataGroup:2024cfk,Tan:2021ahl,Riley:2021pdl, Miller:2021qha}
\bibliography{apssamp}
\clearpage

\appendix
\section{Motivation for percolation}
\label{appA}

In the Introduction, we suggested that we expect exotic matter to appear in the cores of neutron stars. To get a general idea about deconfinement, in particular, we can do a simple estimation to see why this is important. We follow the standard argument suggested by e.g., [122-125], and assume that we only have nucleons (protons and neutrons) and that they occupy a hard volume, V. To provide some numbers, we estimate the size of nucleons to have a radius of $0.84-0.86$ fm based on the measured proton charge radius, proton magnetic radius and neutron magnetic radius from PDG 2024 [126], so we can compare the estimated number density of one nucleon to the baryon (number) densities probed in the interior of a neutron star. Expressing this as an inequality, we have $n_B < 1/V$.

Using our estimate for the radius of a nucleon and roughly approximating a nucleon as a sphere, the density at which the neutron star is the same density as a nucleon is 0.38-0.40 fm$^{-1}$. Beyond these densities, the nucleons would have to overlap and the description of degrees of freedom as simple baryons would not make sense any further. Specifically, within high-mass neutron star cores, depending on model assumptions, one could reproduce densities two or three times this value (see Fig.~4 of [127] for the case of a first-order phase transition). In reality, even before the nucleons overlap, a more complex description of quark deconfinement is needed, which is exactly what the percolation description we apply in this work is meant to reproduce.

\section{Additional cases and details}

\subsection{Percolation parameters and neutron star properties}
\label{appB1}

\begin{figure*}[t!]
\centering 	
\includegraphics[width=1\textwidth, angle=0]{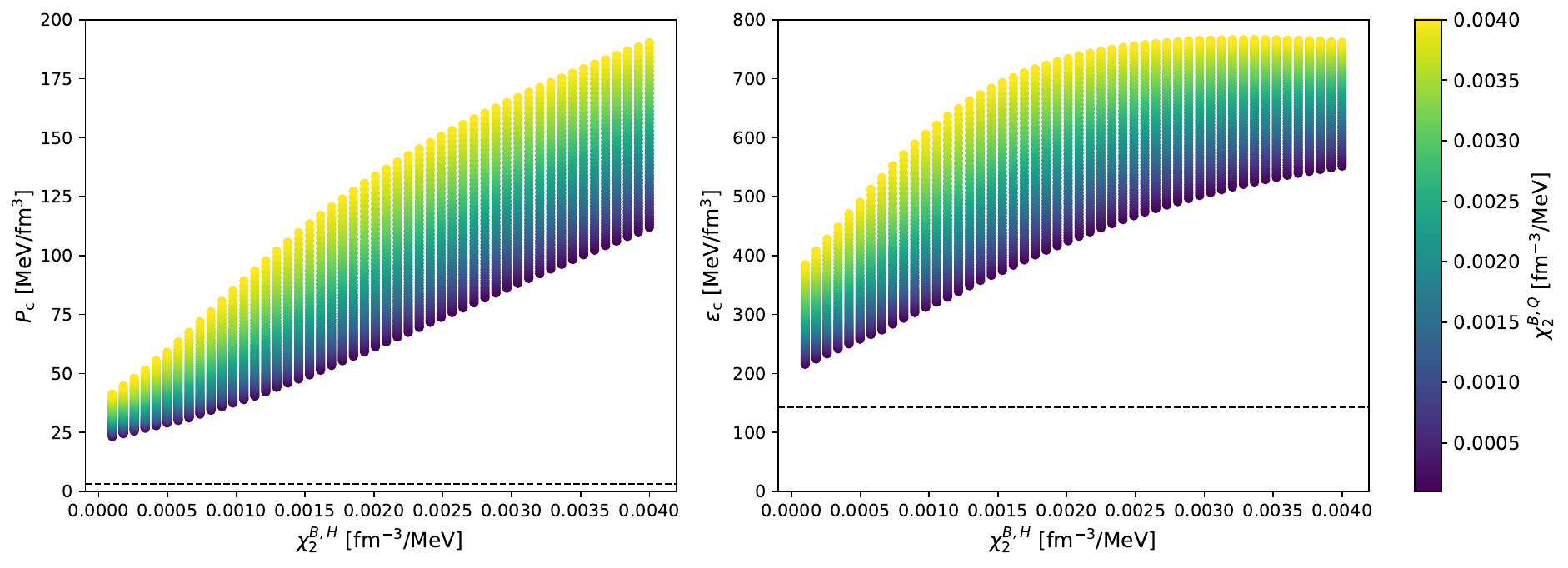}		
\caption{Central stellar pressure (left) and central stellar energy density (right) as functions of the second-order susceptibility at the hadronic boundary of the percolation region. The colored scattering points show the second-order susceptibility at the quark boundary of the percolation region and the black dotted lines show the pressure (left) and energy density (right) at the hadronic boundary. Same percolation region parameters,
density boundaries, and chosen stellar mass used in Fig. 2 for CMF-7.} 
\label{chi2evo}
\end{figure*}

\begin{figure}[t!]
\centering 	\includegraphics[width=0.5\textwidth, angle=0]{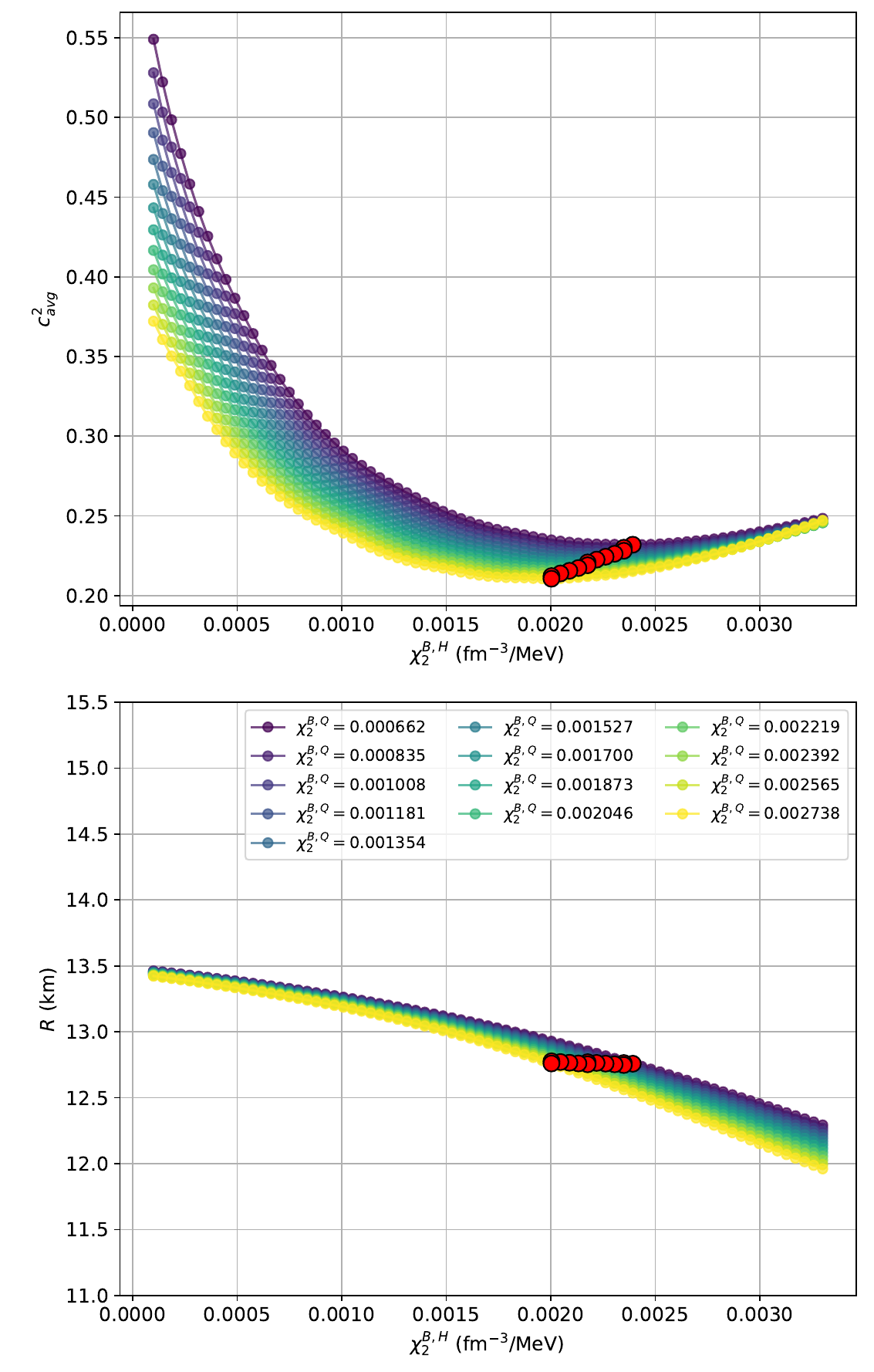}	
\caption{Same as the right panels of Fig. 2 for CMF-7 and a star with mass $1.554\,\rm{M}_{\odot}$, but for different density boundaries for the percolation regions, $n_{B_H} = 0.30$ fm$^{-1}$ and $n_{B_Q} = 1.20$ fm$^{-1}$ (the ones that produce the red curve in Fig. 1.}
\label{v2_fig}
\end{figure}

\begin{figure*}[t!]
\centering 	
\includegraphics[width=1\textwidth, angle=0]{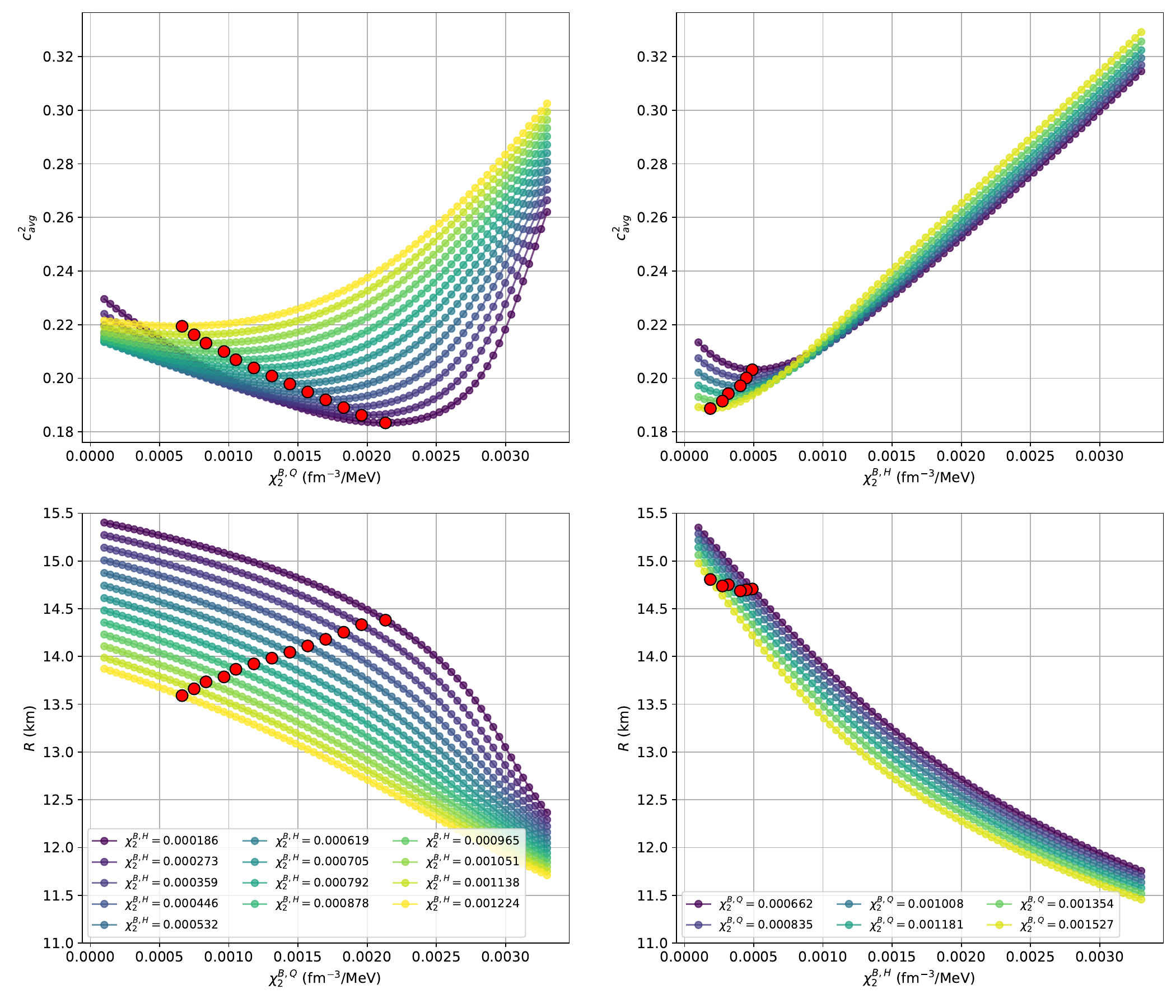}	
\caption{Same percolation region parameters and density boundaries as Fig. 2 for CMF-7, except using a chosen stellar mass of $1.8\,\rm{M}_{\odot}$.} 
\label{1.8mass}
\end{figure*}

\begin{figure*}[t!]
\centering 	
\includegraphics[width=1\textwidth, angle=0]{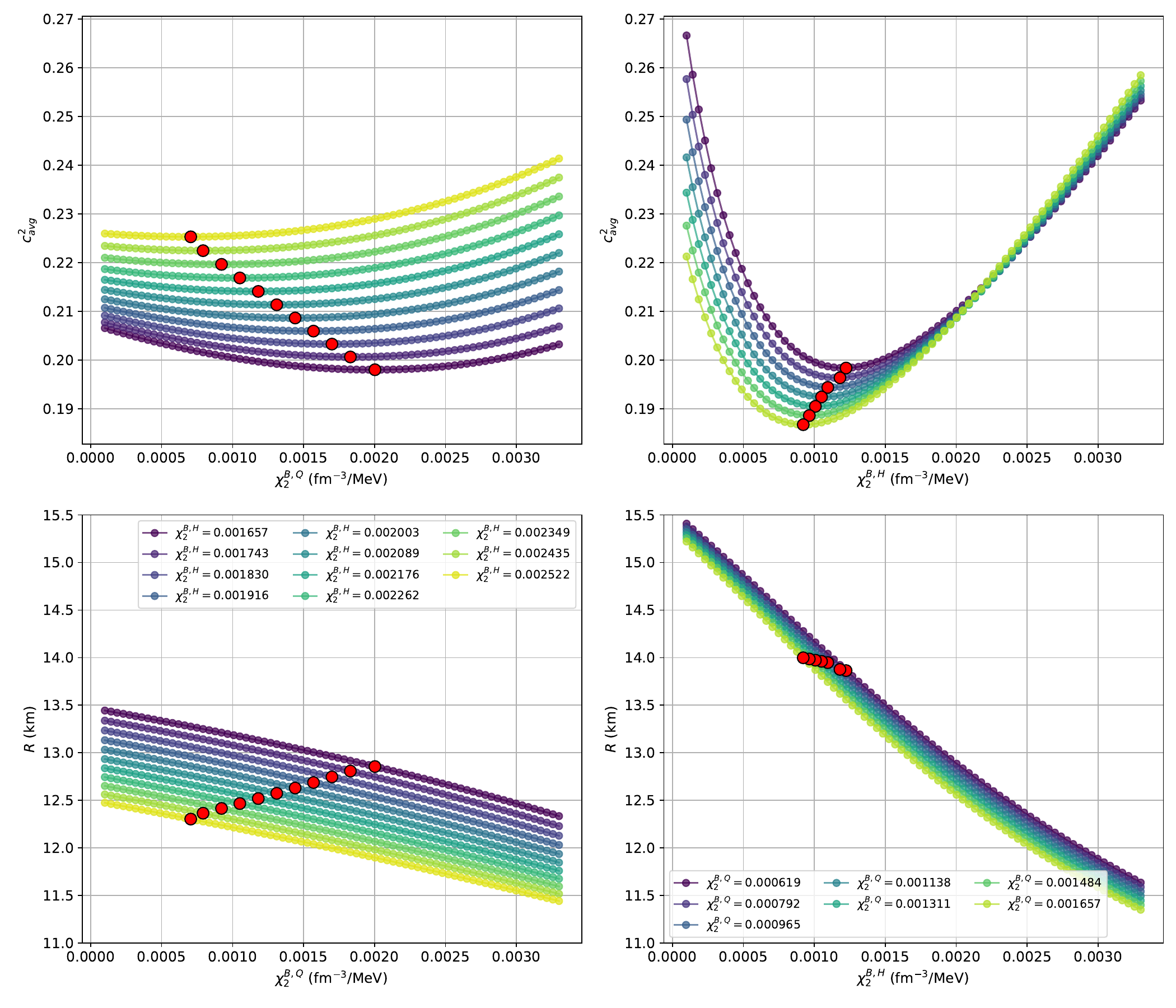}	
\caption{Same percolation region parameters, density boundaries, and chosen stellar mass used in Fig. 2 but using the DD2F-NJL microscopic model.} 
\label{NJL}
\end{figure*}

In this subsection, we illustrate the relationship between the key percolation parameters $\chi_2^{\mathrm{B},\mathrm{H}}$ and $\chi_2^{\mathrm{B},\mathrm{Q}}$ and neutron star properties (for a given stellar mass). Fig.~\ref{chi2evo} shows how both the central stellar pressure, $P_C$, (left panel) and central stellar energy density, $\varepsilon_C$, (right panel) increase with increasing susceptibilities at the boundaries of the percolation region. We observe that $P_c$ increases with both susceptibilities faster than $\varepsilon_c$. This characteristic is what qualitatively explains the appearance of the minimum point of $c^2_{avg}$ we discuss in the text. Quantitatively, the value of $\chi_2^B$ (on either side of the percolation region) corresponding to the minimum of $c^2_{avg}$ (Eq. 3) is determined by
\begin{eqnarray}
\frac{\partial c^2_\mathrm{avg}}{\partial \chi_2^B} = 0\,,
\end{eqnarray}
\begin{eqnarray}
(\varepsilon_c-\varepsilon_H) \frac{\partial P_c}{\partial \chi_2^B} = (P_c-P_H) \frac{\partial \varepsilon_c}{\partial \chi_2^B}\,,
\end{eqnarray}
\begin{eqnarray}
c^2_\mathrm{avg} = \frac{{\partial P_c}/{\partial \chi_2^B}}{{\partial \varepsilon_c}/{\partial \chi_2^B} }\,,
\label{final}
\end{eqnarray}
where the right side of Eq.~\ref{final} is given by the ratio of the slopes of both panels in Fig.~\ref{chi2evo}.

\subsection{Higher density percolation boundary}
\label{appB2}

We now repeat the exact same prescription but change the density boundaries for the percolation regions to match the red curve of Fig. 1 for CMF-7 ($n_{B_H} = 0.30$ fm$^{-1}$ and $n_{B_Q} = 1.20$ fm$^{-1}$), still for a star with mass $1.554\,\rm{M}_{\odot}$.
We once more vary the susceptibility at the quark boundary for the percolation for different values of the susceptibility at the hadronic boundary (the method used in the left panels of Fig. 2), but now obtain no causal EoSs that are minima. We then vary the susceptibility at the hadronic boundary of the percolation for different values for the susceptibility at the quark boundary (the method used in the right panels of Fig. 2). The results are presented in Fig.~\ref{v2_fig}, where the red dots mark minima in the speed of sound squared. The results are qualitatively the same as for the right panels of Fig. 2, but spanning a significantly smaller range of radii of $0.1$ km, still in agreement with the recurring region in the left panels of Fig. 2.

\subsection{Higher mass stars}
\label{appB3}

To further explore the flexibility of our approach, we apply the same prescription described in the text to find a recurring region for a higher chosen stellar mass of $1.8\,\rm{M}_{\odot}$ for CMF-7. We show in Fig.~\ref{1.8mass} recurring regions that appear at different radii in both cases, on the left panels showing the susceptibility at the quark boundary for the percolation in the x-axis and the different susceptibilities at the hadronic boundary using different line colors and on the right panels showing the susceptibilities exchanged. The lower panels show that the radii are now larger (than for the smaller star mass already discussed) in both panels. The radii in the right panel are too large in comparison with astrophysical data, but the radii in the left bottom panel for such a high mass are still reasonable [128, 129].

\subsection{Different microscopic model}
\label{appB4}

The phenomenon of recurring regions we discuss in this work is not unique to the CMF model. We show in Fig.~\ref{NJL} that we can also obtain recurring regions using the DD2F-NJL model, again for an example mass of $1.554\,\rm{M}_{\odot}$. The percolation densities are the same as in Fig. 2 ($n_{B,H}=0.15$ and $n_{B,Q}=1.2$ fm$^{-3}$). The radius for the recurring region occurs at different values, but still not too different from  the ones in CMF-7. Furthermore, all the curves in Fig. 2 and Fig.~\ref{NJL} have the same qualitative shape. We note however that the DD2F-NJL EoSs present a softer hadronic component, which has the effect of lowering the radius of the recurring region in the left panels but having little effect on the right panels.  These results show that using different microscopic models provide a way to tune the location of recurring regions, not only in mass, but also in radii and tidal deformability (related to radius [114, 115]).

\section{Additional plots related to gravitational waves}
\label{appC}

\begin{figure}[t!]
\centering 	
\includegraphics[width=.5\textwidth, angle=0]{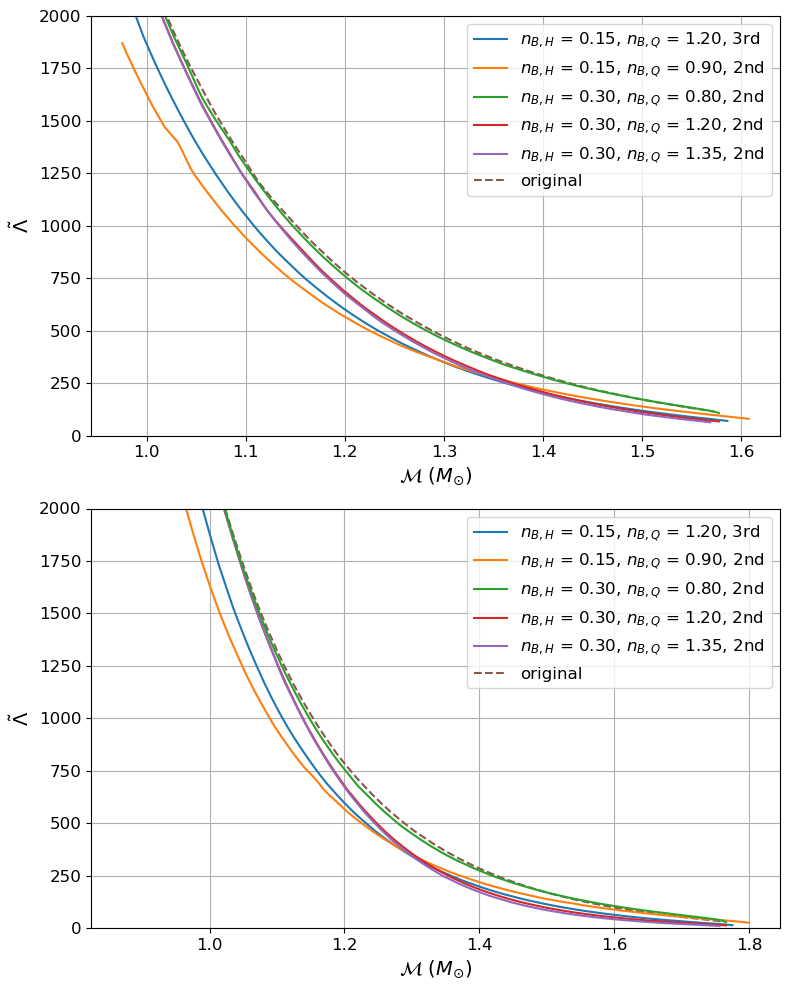}	
\caption{Binary tidal deformability vs. chirp mass for the EoSs shown in Fig. 1. The binary mass ratio is assumed to be 0.8 (upper panel) or 1 (lower panel).} 
\label{chirpmass}
\end{figure}

\begin{figure}[t!]
\centering 	
\includegraphics[width=0.5\textwidth, angle=0]{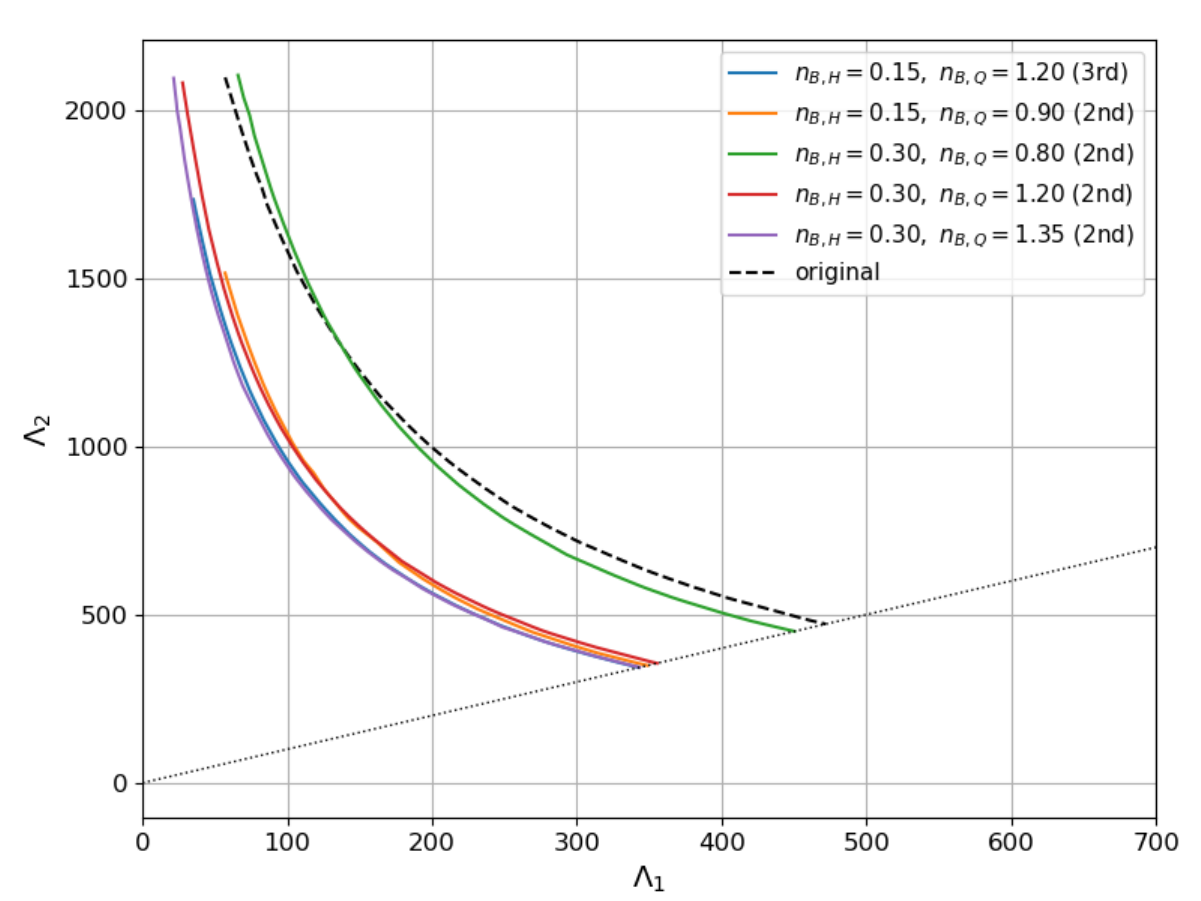}	
\caption{Tidal deformability of two stars in a binary with constant chirp mass $\mathcal{M} = 1.300\,\rm{M}_{\odot}$ for the EoSs shown in Fig. 1.}
\label{lamba1}
\end{figure}

We note that in gravitational-wave detections of binary neutron-star mergers the dimensionless tidal deformability is not directly measured. We show in Fig.~\ref{chirpmass} the binary tidal deformability, $\tilde{\Lambda}$, versus the chirp mass, $\mathcal{M}$, for two different mass ratios, $q$. These quantities are what can be extracted from the observation of gravitational wave events for binary neutron-star mergers. We see the same recurring region appearing that corresponds to those in Fig. 1. This is verified for two different binary mass ratios, $q=0.8$ (top panel) and $q=1$ (bottom panel). 
In addition, in Fig.~\ref{lamba1}, we show $\Lambda_2$ versus $\Lambda_1$ for two stars in a binary merger event at constant chirp mass $\mathcal{M} = 1.300\,\rm{M}_{\odot}$, which is approximately the $\mathcal{M}$  at the recurring region in \ref{chirpmass}. At this chirp mass, the EoSs that pass through the recurring region, nearly overlap for all combinations of  $\Lambda_1$ and $\Lambda_2$, which implies that this recurring region is independent of the mass ratio in the binary.

\section{Summary of steps to obtain a recurring region}
\label{appD}

To produce EoSs that correspond to a mass-radius observation of a neutron star or a mass-tidal deformability observation of a neutron-star merger using our prescription, one would need to
\begin{itemize}
    \item choose a specific stellar mass (maybe the midpoint of the observation data provided) and choose a hadronic EoS and a quark EoS (or one complete one) with output in CompOSE format [90-92]. The EoSs we use, CMF-7~\footnote{\url{https://compose.obspm.fr/eos/273}} and RDF~\footnote{\url{https://compose.obspm.fr/eos/170}}, are available in CompOSE
    \item for chosen densities for the boundaries of the percolation and one susceptibility, run our open source Python scripts~\footnote{\url{https://github.com/np3m/Percolation-Project}} varying the susceptibility at the other percolation region boundary
    \item find the minima of the average speed of sound to determine the radius and tidal deformability of recurring region
    \item change the hadronic EoS (see discussion in the text for why not the quark one) and repeat the process until radius and tidal deformability match desired values
    \item (if desired) vary the chosen densities for the boundaries of the percolation (independently) and the other susceptibility and, for each set, run our Python scripts varying the susceptibility at the other percolation region boundary to find the minima of the average speed of sound to produce more EoSs passing by the same recurring region
\end{itemize}

\end{document}